\documentclass[%
 reprint,
superscriptaddress,
showpacs,
 amsmath,amssymb,
 pre,
]{revtex4-1}

\usepackage{xcolor}
\usepackage{graphicx}% Include figure files
\usepackage{dcolumn}% Align table columns on decimal point
\usepackage{bm}% bold math
\usepackage{nicefrac}
\usepackage{units}

\usepackage[]{natbib} % Use the natbib reference package - read up on this to edit the reference style; if you want text (e.g. Smith et al., 2012) for the in-text references (instead of numbers), remove 'numbers'

\begin{document}

\preprint{APS/123-QED}

\thanks{A footnote to the article title}%

\title{Vesicle Dynamics in a Confined Poiseuille Flow: From Steady-State to Chaos}

\author{Othmane \surname{Aouane}}
\affiliation{Experimental Physics, Saarland University, 66123 Saarbr\"{u}cken, Germany}
\affiliation{Universit\'{e} Grenoble Alpes, LIPHY, F-38000 Grenoble, France}
\affiliation{LMPHE, URAC 12, Facult\'{e} des Sciences, Universit\'{e} Mohammed V- Agdal, Rabat, Morocco}
\author{Marine \surname{Thi\'{e}baud}}
\affiliation{Universit\'{e} Grenoble Alpes, LIPHY, F-38000 Grenoble, France}
\author{Abdelilah \surname{Benyoussef}}
\affiliation{LMPHE, URAC 12, Facult\'{e} des Sciences, Universit\'{e} Mohammed V- Agdal, Rabat, Morocco}
\author{Christian \surname{Wagner}}
\affiliation{Experimental Physics, Saarland University, 66123 Saarbr\"{u}cken, Germany}
\author{Chaouqi \surname{Misbah}}
\affiliation{Universit\'{e} Grenoble Alpes, LIPHY, F-38000 Grenoble, France}
\email{chaouqi.misbah@ujf-grenoble.fr}

\date{\today}% It is always \today, today,
             %  but any date may be explicitly specified

\begin{abstract}
Red blood cells (RBCs) are  the major component of  blood and the flow of blood is dictated by that of RBCs. %
We employ vesicles, which consist of closed bilayer membranes enclosing a fluid, as a model system to study the behavior of RBCs under a confined Poiseuille flow. %
We extensively explore  two main parameters: i) the degree of confinement of vesicles within the channel, and ii) the flow strength. %
Rich and complex dynamics for vesicles are revealed ranging from  steady-state shapes (in the form of parachute and slipper)  to chaotic dynamics of shape. %
Chaos occurs through a cascade of multiple periodic oscillations of the vesicle shape. %
We summarize our  results in a phase diagram in the parameter plane (degree of confinement, flow strength). This finding highlights the level of complexity of a flowing vesicle in the small Reynolds number  where the flow is laminar in the absence of vesicles and can be rendered turbulent due to elasticity of vesicles.%

\end{abstract}

% \pacs{87.16.D-, 83.80 Lz, 47.11.Hj , 47.52.+j}
\maketitle

\section{\label{sec:intro}Introduction}

Nowadays, vesicles are extensively used as a model for understanding  dynamics and deformation of red blood cells (RBCs) at the individual level but also regarding collective phenomena and rheology. %
Vesicle membrane withstands to bending but does not have a shear resistance, unlike RBCs, but they still share several dynamical properties with RBCs, %
like tank-treading and tumbling under linear shear flow, or parachute and slipper shapes under Poiseuille flow \cite{Abkarian08_shear,Petia_review2009,Petia_review2013}. %

Under a Poiseuille flow, the  situation of interest in this paper, it is known experimentally that RBCs exhibit a parachute as well a slipper shape \cite{Skalak69,Schoenbein81,Abkarian08_poiseuille,Tomaiuolo09}. %
Secomb and %
Skalak \cite{Secomb82} have presented a model for the slipper shape based on a lubrication approximation. The slipper shape was also later observed in numerical %
simulations by Pozrikidis \cite{Pozrikidis2005}. These authors used a capsule as a %
model for RBC. Capsules are shells made of polymers and are endowed %
with elastic properties, namely the shear elasticity that mimics the %
RBCs cytoskeleton, i.e. the spectrin network lying underneath the cell membrane. %
More recently, the minimal ingredients for the occurrence of %
a slipper shape were identified \cite{Kaoui09}: %
a two dimensional vesicle even in an unbounded %
Poiseuille flow exhibits a slipper solution when the flow %
strength is comparable to that in the microvasculature. %
The slipper solution occurs as a result of loss of stability of the symmetric %
solution (called also parachute).
These shapes were further investigated by including the effect of quasi-rigid bounding walls \cite{Kaoui11}. %
This study revealed  large variety of shapes and dynamics such as the %
centered and off-centered periodic oscillations (called snaking). These oscillations are regular and stable in time. %
Subsequent study in 3D has also reported on similar phenomena \cite{Fedosov2013,Farutin2014}.
The present study is a follow-up study to that of Kaoui et al.\cite{Kaoui11} and reveals a variety of new states. For example, we find that vesicles can first undergo snaking (periodic oscillation of the shape %
in the form of a snake motion) and suddenly undergoes a new bifurcation showing period-doubling of the temporal oscillation upon variation of a control parameter (e.g. degree of confinement). %
On further variation of control parameter the system undergoes a subharmonic cascade oscillation before transiting to chaos. Other scenarios than period-doubling can also occur as we shall show. %
We investigate the occurrence of chaos using tools of dynamical systems. We present a full phase diagram in parameter space showing variety of dynamics.

\section{\label{sec:model}Theoretical Framework}

 \subsection{Membrane Model}
Vesicles in which we are interested  consist of a closed bilayer fluid membrane.
Typically, vesicles diameter range  from a few %
hundred nanometers to a few hundred micrometers, whereas the thickness of the bilayer %
is around few nm. At room temperature, bilayer membranes may be %
regarded as two-dimensional fluids, but one should keep in mind that they may present %
other phases (e.g., crystal and solid or gel phases) depending on the temperature and the %
chemical nature of the lipids. The membrane is a two-dimensional %
incompressible fluid, therefore its area  is locally conserved. %
Due to membrane  impermeability,  the volume  of the enclosed liquid inside the vesicle is also conserved. %
Fluid membranes present a viscous resistance to shear stress, leading to a deformation %
of the membrane with no storage of elastic energy.
The only energetic contribution comes from the bending energy. Here we employ the Helfrich elasticity
theory for bilayer membranes to describe the curvature energy in 2D \cite{Helfrich73} (2D models have proven to capture the essential features of vesicles under flow, and will be adopted here),
\begin{equation}
\label{eq::Hamiltonian}
 \mathcal{H}  = \frac{\kappa}{2} \oint({c - c_0})^2 ds %
\end{equation}
where $c$ and $c_0$ are respectively the mean and spontaneous curvatures, $\kappa$ is the curvature elastic modulus, $s$ denotes the curvilinear coordinate along
the membrane, and $\oint$ refers to an integral over the (2D) membrane contour. %
In two dimensions the spontaneous curvature is irrelevant owing to the property $\oint{c} ds=2\pi$ (an irrelevant constant).
A tension-like energy is added to the bending energy \eqref{eq::Hamiltonian} in order  to fulfill the vesicle local perimeter conservation constraint.
\begin{equation}
 E_{tens} =  \oint \zeta (s) ds
\end{equation}
where $\zeta$ is a Lagrange multiplier that enforces constant local length.
The force is obtained from  the functional derivative of the Hamiltonian, including the tension energy, with the respect to the membrane elementary displacement \cite{Kaoui08}. %
		\begin{equation}
		 \mathbf{F}_{mem} =  \kappa [\frac{\partial^2{c}}{\partial s^2} + \frac{c^3}{2}]\mathbf{n}  - c \zeta \mathbf{n} +
		 \frac{\partial{\zeta}}{\partial{s}}\mathbf{t}
		\end{equation}
where $\mathbf{n}$ and $\mathbf{t}$  are the normal and tangent unit vectors, respectively.   			
The vesicle is characterized by a reduced area ($\tau$), combining the actual fluid area enclosed by the vesicle contour  $(S = \pi R_0^{2})$ and the area of a disk having the same perimeter as the vesicle with %
\begin{equation}
 \tau =  \frac{2\sqrt{S\pi}}{P}
\end{equation}
and a viscosity contrast ($\lambda$) which expresses the ratio between the inner ($\eta_{in}$) and outer fluid ($\eta_{out}$) viscosities
\begin{equation}
 \lambda = \frac{\eta_{in}}{\eta_{out}}
\end{equation}
The effective radius of the cell $(R_{0}\equiv \sqrt{S/\pi})$ and the outer viscosity $(\eta_{out})$ are chosen to be the characteristic length and viscosity scales, respectively.
	
 \subsection{Boundary Integral Formulation}
The boundary integral method for low Reynolds number flow is well established \cite{Pozrikidis92}, and we have used it in different
contexts in 2D and 3D  \cite{Cantat99,Kaoui08,Kaoui09,Selmi11,Biben11,Marine13,Farutin2014}. Here we shall use a special Green function introduced quite recently in \cite{Marine13} that automatically satisfies the no-slip boundary condition at the bounding walls.
Using this special Green function the velocity along the membrane is given by \cite{Marine13}
\begin{widetext}
\begin{equation}
	\label{eq::velocity_integral}
	\frac{1+\lambda}{2} {\mathbf{v}}(\mathbf{X}) =  {\mathbf{v}}^{\infty}(\mathbf{X}) +%
	 \frac{1}{{\eta}_{out}}\int_{\gamma}{\mathbf{G}_w}(\mathbf{X}-\mathbf{X}_{0}){\mathbf{F}_{mem}}(\mathbf{X}_{0})ds({\mathbf{X}_{0}}) + %
	(1-\lambda) \int_{\gamma}\mathbf{v}(\mathbf{X}_{0}){\mathbf{K}_w}(\mathbf{X}-\mathbf{X}_{0}){\mathbf{n}} %
	(\mathbf{X}_{0})ds({\mathbf{X}_{0}}) %
\end{equation}
\end{widetext}
where $\mathbf{X}$ and $\mathbf{X}_{0}$ are two position vectors belonging to the membrane ($\gamma$). $\mathbf{v}$ and ${\mathbf{v}}^{\infty}$ are the membrane's velocity, and the imposed velocity. ${\mathbf{G}_w}$ and ${\mathbf{K}_w}$ stand for the Green second and third order tensors for two parallel walls, and ${\mathbf{v}}^{\infty}(\mathbf{X})$ is the imposed Poiseuille flow imposed (to be specified below).
The detailed expression of the Green functions is given in Ref. \cite{Marine13}. Because of the use of this special Green function the integral is only performed along the vesicle, and not along the bounding walls. This provides us with a quite powerful technique, as recently demonstrated \cite{Marine13,Marine14}.
The boundary conditions used in order to derive the integral equation (\ref{eq::velocity_integral})  are: i) no-slip condition at the walls and at the membrane, ii) stress balance at the membrane, and iii) membrane inextensibility.
	
The external flow and confinement introduce two additional dimensionless numbers: the so-called capillary number ($C_k$) to quantify the flow strength over bending forces, and the confinement ($C_n)$ to describe the ratio between the effective diameter of the vesicle and the width of the channel.
The imposed Poiseuille flow is  written as
	\begin{equation}
		 \left\{
		  \begin{array}{rcr}
		    {v_x}^\infty & = & v_{max}[1-(\frac{y}{W/2})^2] \\
		    {v_y}^\infty & = & 0 \\
		  \end{array}
		\right. 			
	\end{equation}

The capillary number is defined as
	\begin{equation}
		C_k = \frac{\eta_{out} R_{0}^4}{\kappa}\frac{v_{max}}{(W/2)^2}\equiv \tau_c \dot\gamma
	\end{equation}
and the confinement as
	\begin{equation}
		C_n = \frac{2 R_0}{W}
	\end{equation}
where $R_0$, $W$ and $v_{max}$ are the effective radius of the cell, the width of the channel and the maximum velocity of the unperturbed Poiseuille flow. %
We define the characteristic shear rate $\dot\gamma$ as the imposed velocity gradient evaluated at $y=R_0/2$, %
and it is equal to ${R_0}v_{max}/{(W/2)^2}$, and $\tau_c=\eta_0 R_0^3/\kappa$ is the characteristic shape relaxation time. Time will be measured hereafter in unit of $\tau_c$ and distances in unit of $R_0$.
The details of numerical treatments are similar to those used in Refs.  \cite{Ghigliotti2010,Biben11}.
%%%%%%%%%%%%%%%%%%%%%%%%%%%%%%%%%%%%%%%%%%%%%%%%%%%%%%%%%%%%%%%
%%%%%%%%%%%%%%%%%%%%%%%%%%%%%%%%%%%%%%%%%%%%%%%%%%%%%%%%%%%%%%%
\section{\label{sec:results}Results and Discussion}
We performed a systematic scan in the three  dimensional  parameter space ($\lambda$, $C_k$, $C_n$), in order to explore the various intricate behaviors  of a  vesicle  under a Poiseuille flow. %
Instead of $C_k$ we shall use the combination $C_k W/R_0 = C_k/2C_n$, which corresponds to the definition of capillary number of \cite{Kaoui11,Tahiri12}. This will simplify comparison with the results of \cite{Tahiri12}. %
In all simulations, we have set the reduced area $\tau$ to $0.6$ which is close to the one of a rbc in 2D.%

\subsection{Effect of flow strength and confinement on the shape of a vesicle (case $\lambda = 1$)}
We first set viscosity contrast to $\lambda=1$  and  explored %
the effect of the confinement and the capillary number on the morphology of the cell. %
In order to test the new code based on the Green's function that vanishes at the wall \cite{Marine13}, %
we have first confirmed the previously  reported results \cite{Kaoui11,Tahiri12}, namely the existence of six different states: parachutelike shape, %
the confined and unconfined slipperlike shape, the centered and off-centered oscillating motion (called snaking in \cite{Kaoui11}) and peanut-like shape \cite{Tahiri12}). %
Fig.~\ref{fig::ampli_time_para_slipper} shows the parachutelike and confined slipperlike solutions.
The snaking motions (centered and non centered) recently reported by  Kaoui et al. \cite{Kaoui11} and Tahiri et al. \cite{Tahiri12}, have not exhausted all intricate dynamics. %
By investigating the evolution of solutions under close scrutiny we have discovered a variety of new states ranging from simple oscillations to complex multi-periodic oscillations, until chaotic motion prevails, as described below. %
\subsubsection{Transition to chaos via a subharmonic cascade:}
\begin{figure}[h!]
   \begin{minipage}[c]{0.45\textwidth}		
      \centering	
      \includegraphics[width = 1.0 \linewidth]{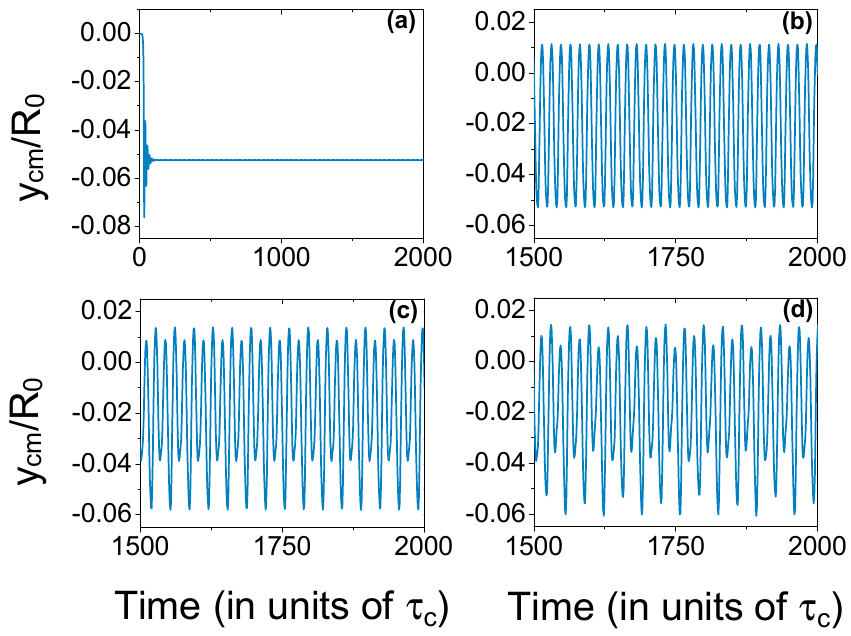}
   \end{minipage}
    \caption{(Color online) The center of mass vertical position as a function of time. (a) slipper solution ($C_k W/R_0= 5$ and $C_n=0.769$); (b) snaking ($C_k W/R_0= 5$ and $C_n = 0.733$); (c) period-doubling ($C_k W/R_0= 5$ and $C_n = 0.729$); %
    and (d) period-quadrupling dynamics ($C_k W/R_0= 5$ and $C_n=0.727$).}
    \label{fig::ampl_time}						
\end{figure}

\begin{figure}[h!]
   \begin{minipage}[c]{0.45\textwidth}	
    \includegraphics[width = 1.0 \linewidth]{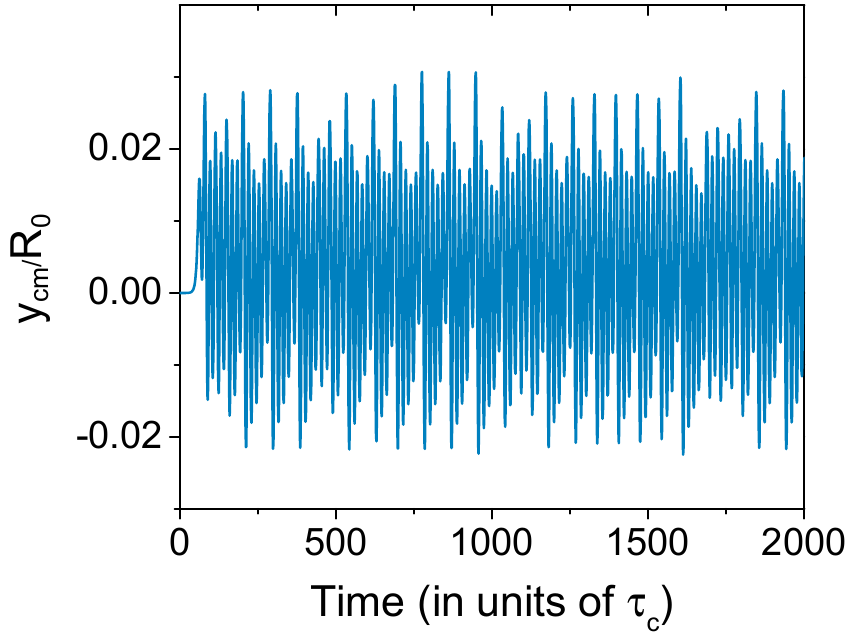}
   \end{minipage}
    \caption{(Color online) The center of mass vertical position as a function of time. An apparently chaos regime is found for $C_k W/R_0=5$ and $C_n=0.689$. }
    \label{chaos}						
\end{figure}

\begin{figure}[h!]
   \begin{minipage}[c]{0.45\textwidth}	
    \centering		
    \includegraphics[width = 1.0 \linewidth]{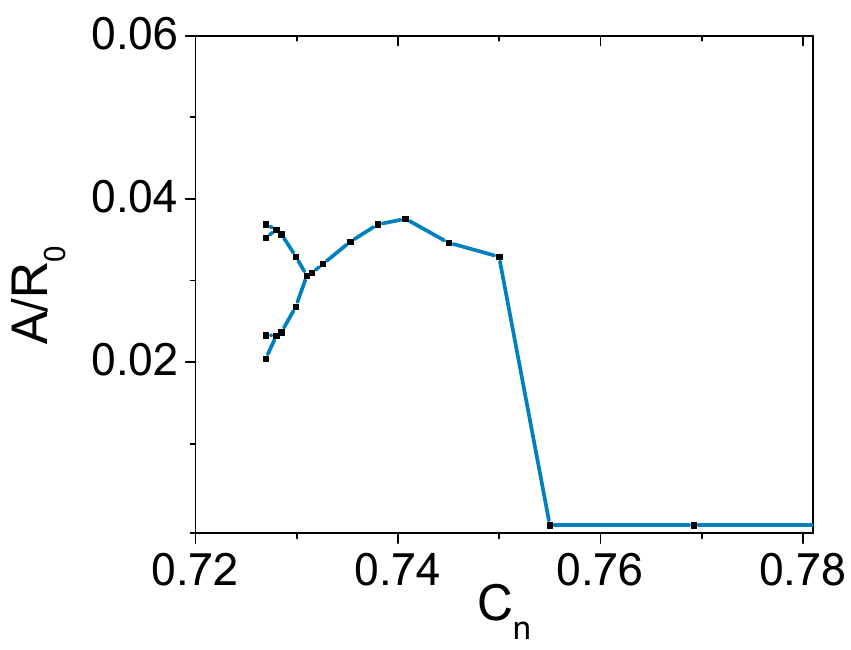}
   \end{minipage}
    \caption{(Color online) Period-doubling bifurcation diagram. The capillary number is fixed ($C_k W/R_0=5$), and only the confinement $C_n$ is changed. The $x$ and $y$ axis stand for the confinement and the amplitude of the oscillations.}
    \label{fig::Bifurcation_Diagram2}						
\end{figure}

We have set $C_k W/R_0=5$ and varied the degree of confinement $C_n$. The results are shown in (Fig.\ref{fig::ampl_time}) where we represent the vertical position of vesicle center of mass ($y_{cm}$) as a function of time. %(when the vesicle has reached a permanent regime, i.e. we ignore transient motions).
Below a first critical value of $C_n$, the slipper becomes unstable in favor of a snaking motion (off-centered). This is a Hopf bifurcation. %
Close to bifurcation point the temporal evolution of the  amplitude of lateral excursion of center of mass (${y}_{cm}$) remains constant over time %
(see Fig. \ref{fig::ampl_time}b). By reducing further $C_n$ the simple snaking solution undergoes a new bifurcation whereby %
the period of oscillation has doubled (Fig. \ref{fig::ampl_time}c) and then quadrupled for a smaller value of $C_n$ (Fig. \ref{fig::ampl_time}d). %
By decreasing further $C_n$ dynamics enter a chaotic regime (Fig. \ref{chaos}).
\begin{figure}[h!]
   \begin{minipage}[c]{0.45\textwidth}		
      \centering	
      \includegraphics[width = 1.0 \linewidth]{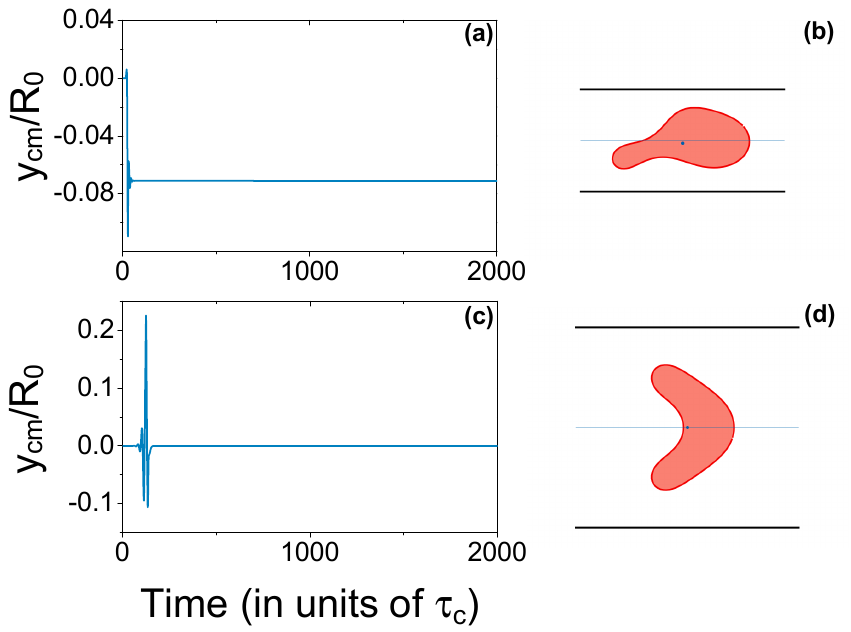}
   \end{minipage}
    \caption{(Color online) Stationary shape and history of the center of mass vertical position as a function of time: (a) and (b): slipper solution ($C_k W/R_0=6.25$ and $C_n=0.8$); (c) and (d): parachute solution ($C_k W/R_0=6.25$ and $C_n=0.44$). %
    The x axis codes for the lateral position of the mass center %
    $y_{cm}$ of the cell scaled by the effective radius of the cell ($R_0$).}
    \label{fig::ampli_time_para_slipper}						
\end{figure}
In Fig.~\ref{fig::Bifurcation_Diagram2}, we represent the amplitude $A$ of excursion of center of mass in the $y$-direction (that is the absolute value of difference between two successive maxima). %
Since a slipper (as well as a parachute solution) moves along a line in the $x$-direction (cf. Fig.\ref{fig::ampli_time_para_slipper}) the amplitude of lateral excursion is zero above a critical value of $C_n = 0.75$ (Fig.~\ref{fig::Bifurcation_Diagram2}). %{\color{red} we should show "amplitude as function of time" for the there motions}.  %
Fig.~\ref{fig::Bifurcation_Diagram2} shows the amplitude as a function of $C_n$, where we can see the beginning of sub-harmonic cascade, %
and the signature of accumulation of bifurcation points. This is a universal behavior, as well documented in chaos textbooks \cite{Schroeder12,Ott02,Berge92}. %
The subharmonic cascade is one of the three generic scenarios of transition to chaos (the two others being intermittency and quasi-periodicity). %
Here we have represented only the main oscillation (period 1), the period doubling (period 2) and quadrupling (period 4).
Because of the universal accumulation  in the sub-harmonic cascade (that is the location points of new bifurcations to higher order oscillations become closer and closer), %
the transition to period-8 and 16 for example requires tuning very carefully the control parameter as well as increasing numerical precision %
(a significant reduction of the numerical mesh size leads to excessive computation time) and it was not our aim to provide a very detailed analysis of the higher order period-doubling cascade. %
Starting from the regime of period-4 oscillation, we found that a quite small variation of $C_n$ (of about $4\%$)  leads to chaos, as shown in Fig. \ref{chaos}. %
\subsubsection{Transition to chaos via a period-tripling bifurcation:}
\begin{figure}[h!]
   \begin{minipage}[c]{0.45\textwidth}	
    \centering		
    \includegraphics[width = 1.0 \linewidth]{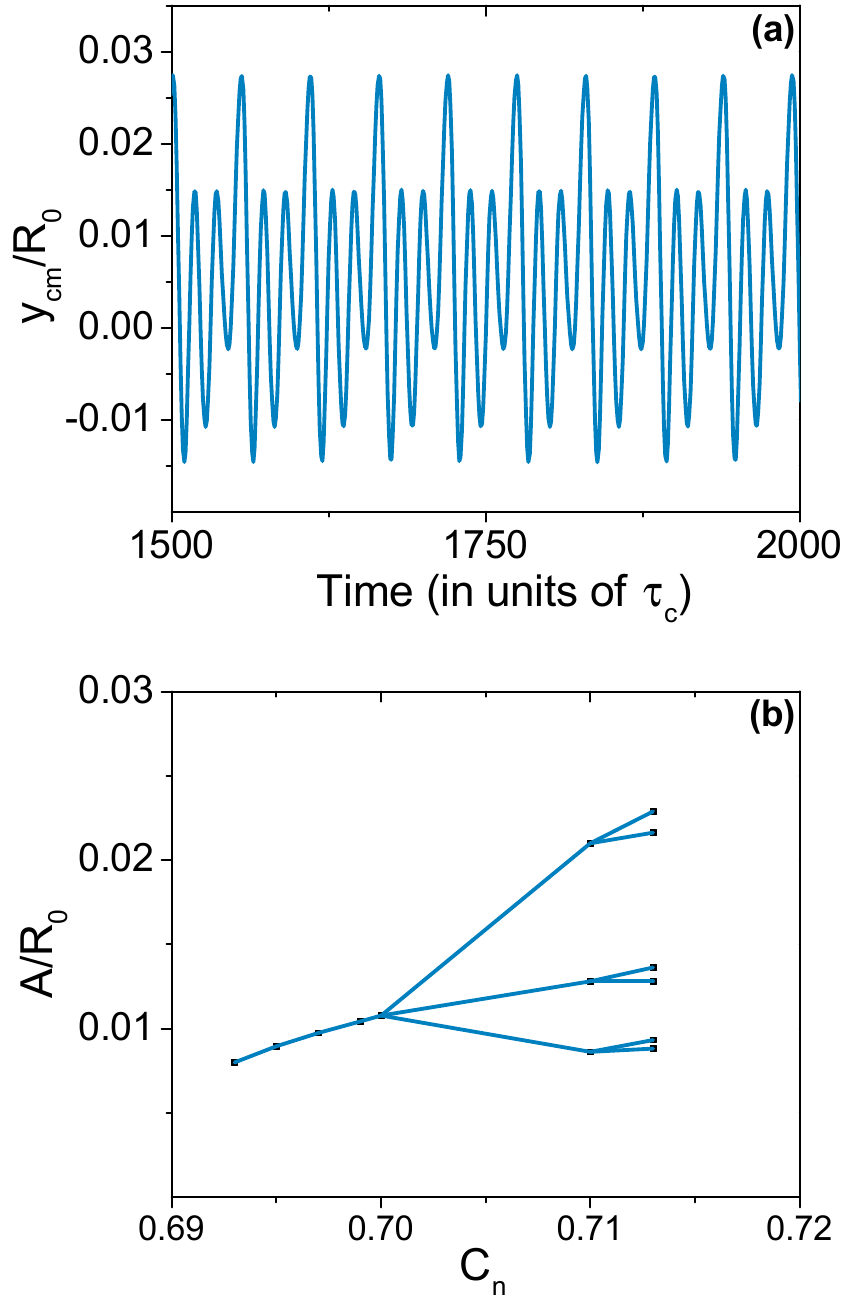}
   \end{minipage}
    \caption{(Color online) Period-tripling motion of the mass center. (a) temporal behavior for a period tripling dynamics ($C_k W/R_0=4.611$ and $C_n=0.71$).  (b) bifurcation diagram ($C_k W/R_0=4.611$).}
    \label{fig::Bifurcation_Diagram3}						
\end{figure}

The subharmonic cascade is one of the three classical scenarios of transition to chaos (in addition to intermittency and quasi-periodicity). %
The sub-harmonic cascade corresponds to a cascade where at each bifurcation point the period is doubled (or the frequency is halved). %
By analyzing the dynamics of the initial snaking motion in other regions of parameter space, we have discovered that the snaking motion can also loose its stability in favor of %
a period-tripling bifurcation, which is a less known  scenario as compared to the period doubling one. %
We show in Fig. \ref{fig::Bifurcation_Diagram3} both a typical temporal signal and the bifurcation diagram. %
Period-tripling bifurcations and more complex transitions were also reported in literature. We take as reference, for instance, the pioneering paper of Li and Yorke \cite{Li75} where %
they introduced the first mathematical definition of discrete chaos, showing the relation between the period three and chaos. %
Lui \cite{Lui13} presented sufficient mathematical conditions for period-tripling and period-n bifurcations. Ze-Hui et al \cite{ze2006subharmonic} reported %
subharmonic bifurcations in a granular system, in the sequence of period-tripling, period-sextupling, and chaos. Zhusubaliyev and Mosekilde \cite{Zhusubaliyev03} showed %
transition from periodic to chaotic oscillations through period-doubling, -tripling, -quadrupling, -quintupling, etc., bifurcations. %
They also discussed more complex transitions, from a family of cycle to another family of cycles with multiple periods. %
\begin{figure}[h!]
      \centering
    \includegraphics[width = 1.0 \linewidth]{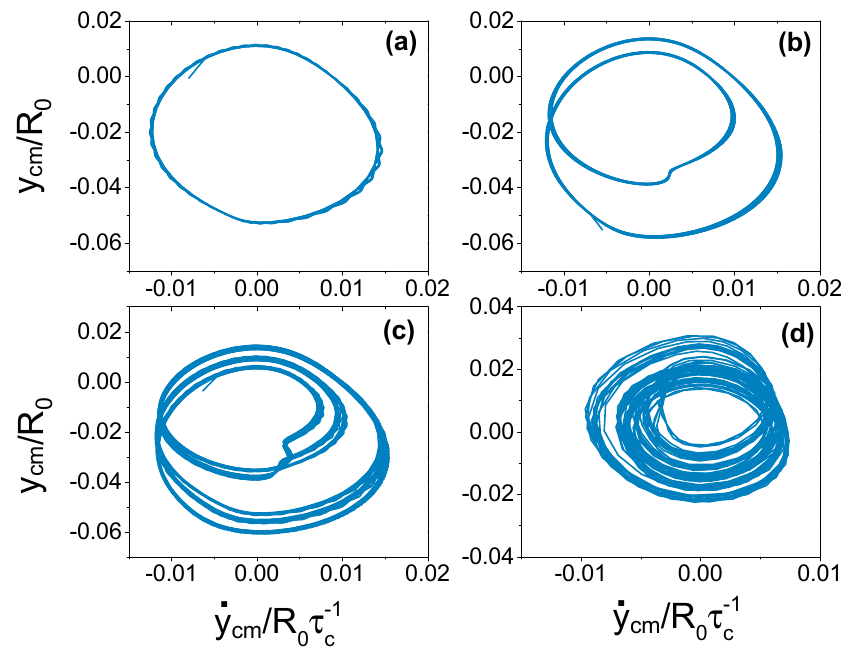}
    \caption{(Color online) Poincar\'{e} sections of different oscillations observed by decreasing the confinement when the capillary number $C_k W/R_0$ has been fixed at 5. The y-axis stands for the y-component of the mass center %
    of the cell and the x-axis for its derivative with respect to time. a)- $C_n = 0.733$. b)- $C_n = 0.729$. %
    c)- $C_n = 0.727$. d)- $C_n = 0.689$.}
    \label{fig::Poincare_sections}						
\end{figure}
\begin{figure}[h!]
   \begin{minipage}[c]{0.45\textwidth}		
      \centering	
      \includegraphics[width = 1.0 \linewidth]{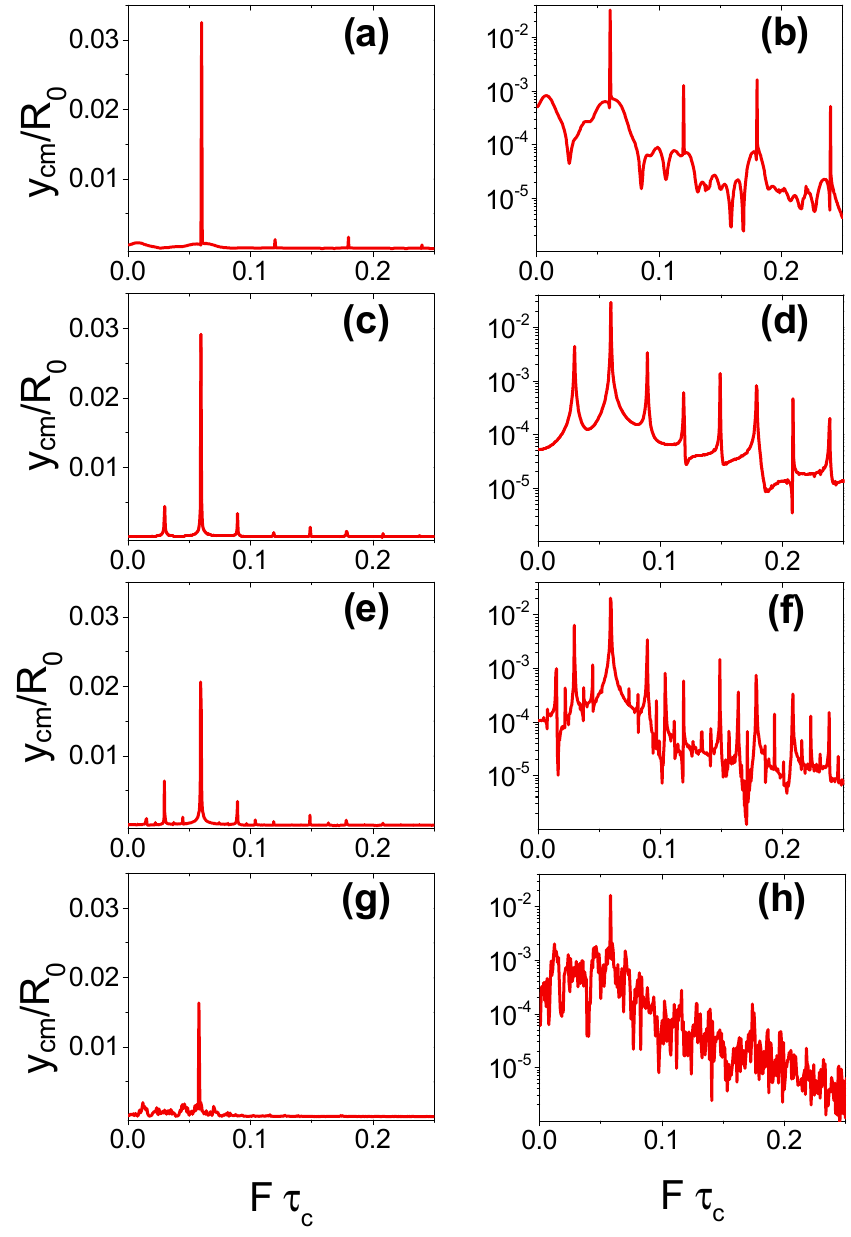}
   \end{minipage}
    \caption{(Color online) Fourier spectrum of different dynamics in linear (a), (c), (e), and (g) and semi-log (b), (d), (f), and (h) coordinates: %
    \textbf{a} and \textbf{b} snaking ($C_k W/R_0 = 5$ and $C_n = 0.733$); \textbf{c} and \textbf{d} period-doubling ($C_k W/R_0 = 5$ and $C_n = 0.729$); \textbf{e} and \textbf{f} period-quadrupling ($C_k W/R_0 = 5$ and $C_n = 0.727$); %
    and \textbf{g} and \textbf{h} chaotic dynamics ($C_k W/R_0 = 5$ and $C_n = 0.689$).  %
    The semi-log scale allows to see more easily the continuum spectrum characteristic of chaotic regimes. %
    The x axis codes for the frequency $F$ scaled by the characteristic time $\tau_c$.}
    \label{fig::fourier}						
\end{figure}
 \begin{figure}[h!]
   \begin{minipage}[c]{0.45\textwidth}	
    \centering
	\includegraphics[width = 1.0 \linewidth]{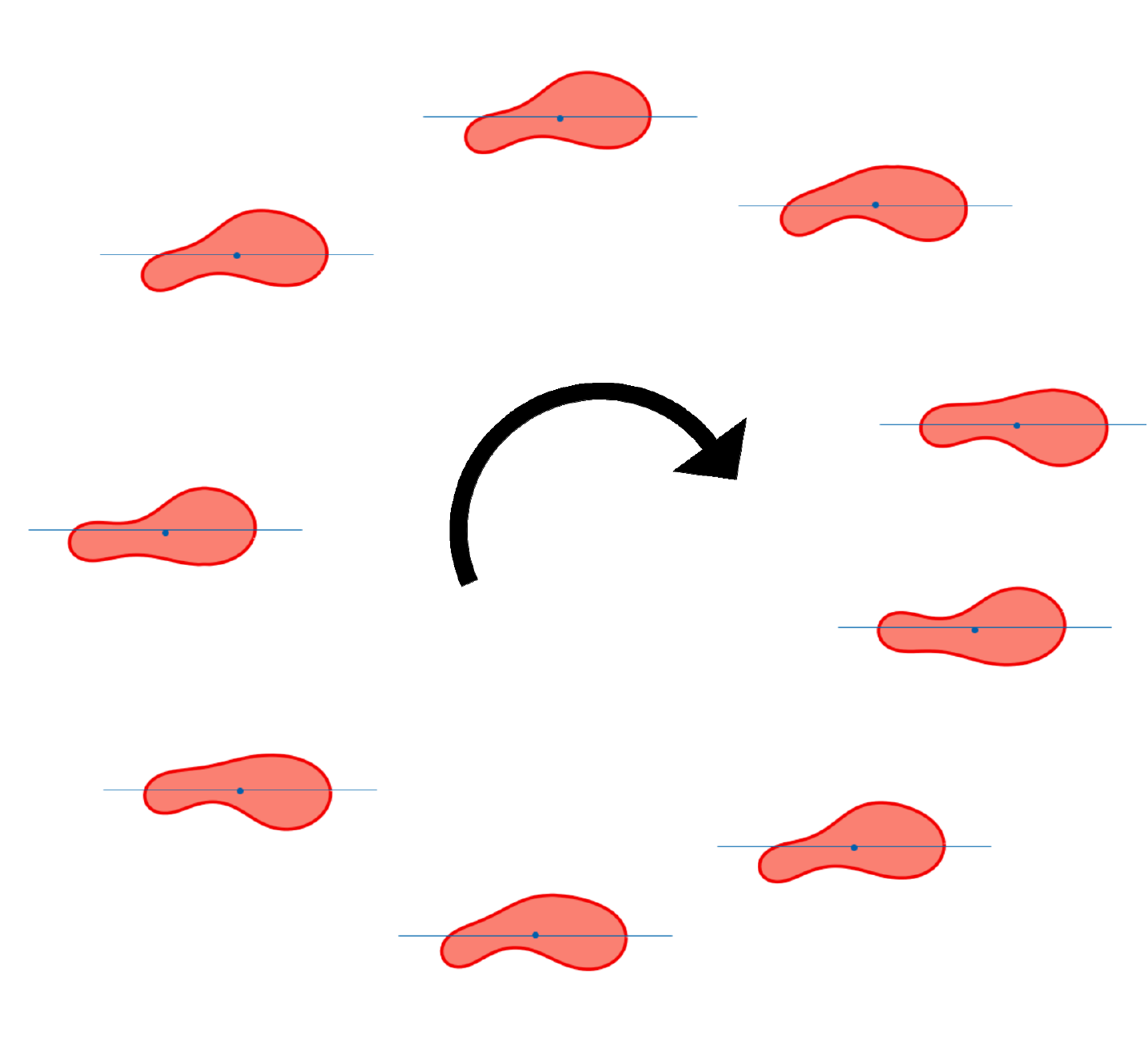}
   \end{minipage} 	
    \caption{(Color online) Snapshots of a cycle of period-doubling dynamics ($C_k W/R_0 = 5$ and $C_n = 0.729$). The cell seems to move like a spermatozoon, using its tail-like as a flagellum. %
    The straight solid blue line indicates the centerline of the channel %
    and the blue point codes for the mass center of the cell.}
    \label{fig::Snap_osci}
\end{figure}
\begin{figure}[h!]
   \begin{minipage}[c]{0.45\textwidth}	
    \centering
    \includegraphics[width = 1.0 \linewidth]{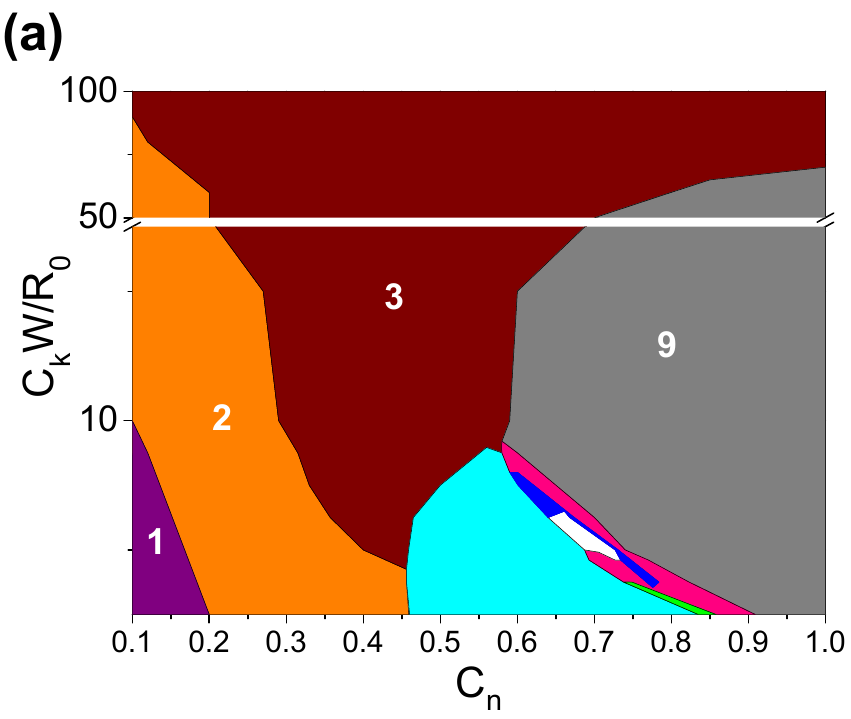}
   \end{minipage}
   \begin{minipage}[c]{0.45\textwidth}	
    \centering
    \includegraphics[width = 1.0 \linewidth]{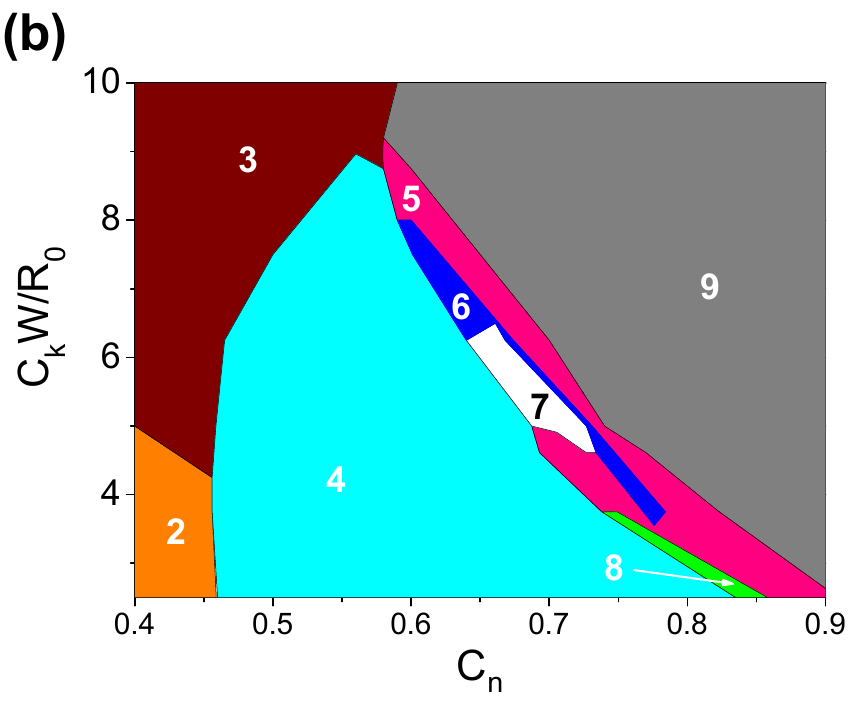}
   \end{minipage}
   \begin{minipage}[c]{0.45\textwidth}	
    \centering
\vspace{0.75cm}
    \includegraphics[width = 1. \linewidth]{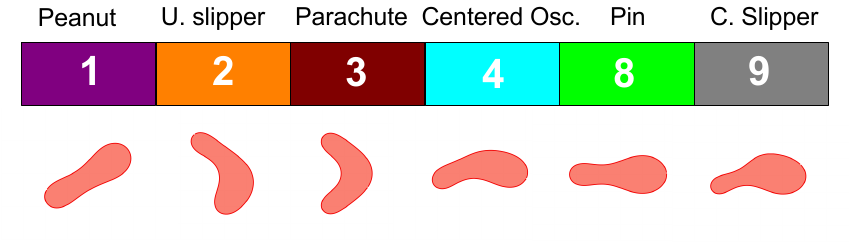}
   \end{minipage}
    \caption{(Color online) Phase diagram summarizing the different morphologies and dynamics of a single vesicle with a viscosity contrast set to unity ($\lambda = 1$). %
    (a) overview of the phase diagram. (b) zoom on the region where oscillations occur. %
    The combined effect of the confinement and flow strength leads to 9 distinct regions represented by different colors in the phase diagram: peanut-like shape (purple); %
    unconfined slipperlike shape (orange); parachutelike shape (dark red); confined slipperlike shape (grey); centered oscillations (cyan); multiple periodic oscillations (dark blue); %
    chaotic oscillations (white); off-centered oscillations (pink); and pin-like shape (green). %
    Note that three regions are not represented in the legend %
    namely: the multiple periodic, chaotic, and off-centered oscillations. %
}
    \label{fig::PhDiL1}
\end{figure}
\subsubsection{Transition to Chaos:}
In order to characterize chaotic dynamics we have performed a Poincar\'e map as well as Fourier transforms of the temporal  evolution of center of mass amplitude.
Fig. \ref{fig::Poincare_sections} displays the Poincar\'{e} sections relative to the different $1$, $2$, and $4$ periodic oscillations, in addition to the chaotic one. %
In this case, the gradual decrease of the confinement under a low capillary number ($C_k W/R_0 =5$ in these simulations) is responsible of the observed transitions. %
The  motion actually can be assimilated to a flagella-like motion, where the vesicle undergoes a periodic or a chaotic up-down motion. %
 Snapshots of this motion are shown in Fig.\ref{fig::Snap_osci} (See Supplemental Material at [\url{SM}] for the different dynamics). Fig. \ref{fig::fourier} %
shows the Fourier spectrum of different dynamics. We see there the occurrence of the cascade until the transition to chaos. %
\subsection{Phase diagram}
We have performed a systematic analysis in order to determine the region of different dynamical manifestation going from order to chaos. %
The results are shown in Fig. \ref{fig::PhDiL1}. Besides the dynamics and shapes  reported earlier \cite{Kaoui11,Tahiri12}, %
revealing slipper, parachute and snaking, we have identified here more complex dynamics, going from higher order oscillatory motion until chaos. %
Surprisingly enough, a simple situation treated here, namely a 2D vesicle under a Poiseuille flow, has revealed broadly 9 different kinds of motion %
(actually the number is even larger, since in Fig. \ref{fig::PhDiL1} we do not specify the kind of multiple oscillation). This result highlights %
the complexity of this free boundary problem, where membrane elasticity that acts here only via bending forces can trigger rich dynamics.

\subsection{RBC-like vesicles in microcirculation conditions}
\begin{figure}[h!]
   \begin{minipage}[c]{0.45\textwidth}
    \centering
    \includegraphics[width = 1.0 \linewidth]{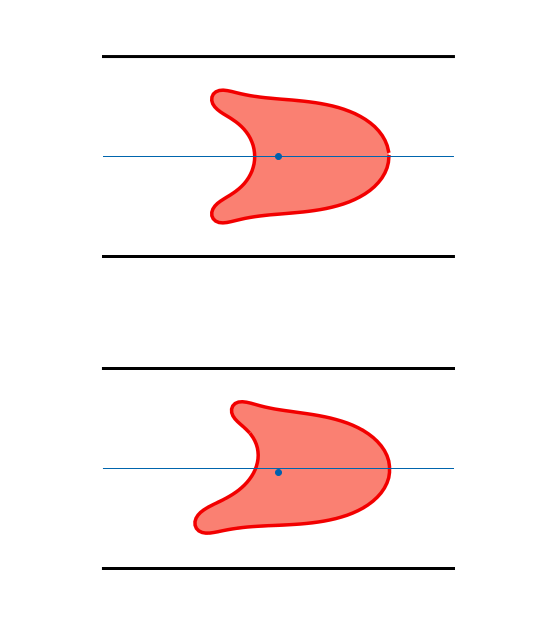}
   \end{minipage}
    \caption{(Color online) Stationary shapes exhibited by a rbc-like vesicle under the same conditions of flow and confinement ($C_k=120$; and $C_n=0.7$). Top: $\lambda = 5$ (cytoplasmic viscosity $\approx$\unit[5]{cP}). %
    Bottom: $\lambda = 10$ (cytoplasmic viscosity $\approx$ \unit[10]{cP}).}
    \label{fig::Shapes_L5-L10}
\end{figure}

\begin{figure}[h!]
   \begin{minipage}[c]{0.45\textwidth}	
    \centering
    \includegraphics[width = 1.0 \linewidth]{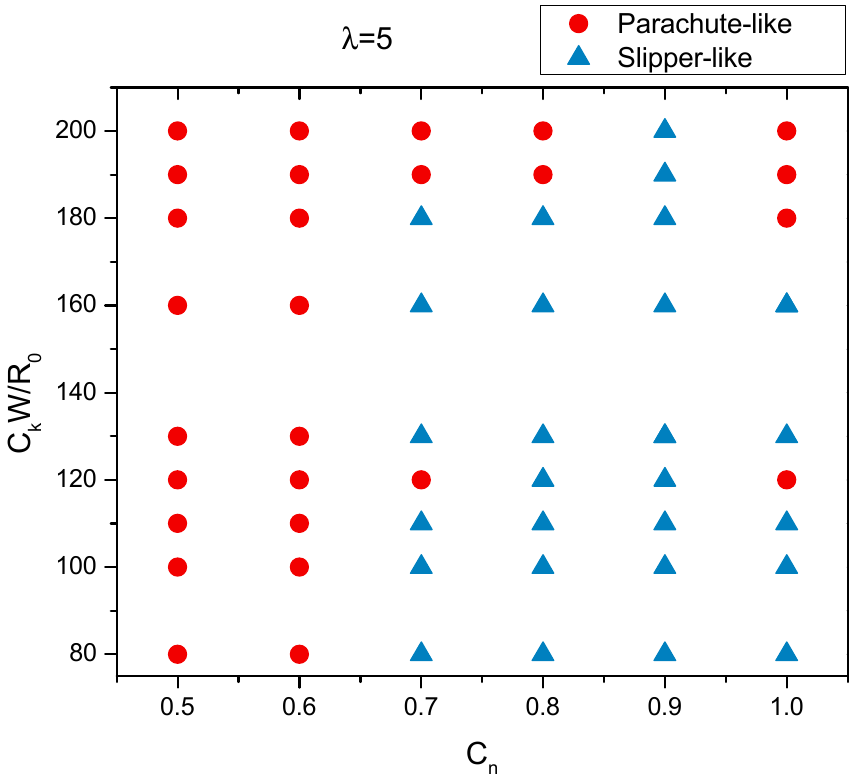}
   \end{minipage}
    \caption{(Color online) Phase diagram of a red cell-like vesicle ($\lambda=5$) in a Poiseuille flow showing the existence of 3 different regimes at very high capillary number.} %
   \label{fig::PhDiL5}
 \end{figure}

\begin{figure}[h!]
   \begin{minipage}[c]{0.45\textwidth}	
    \centering
    \includegraphics[width = 1.0 \linewidth]{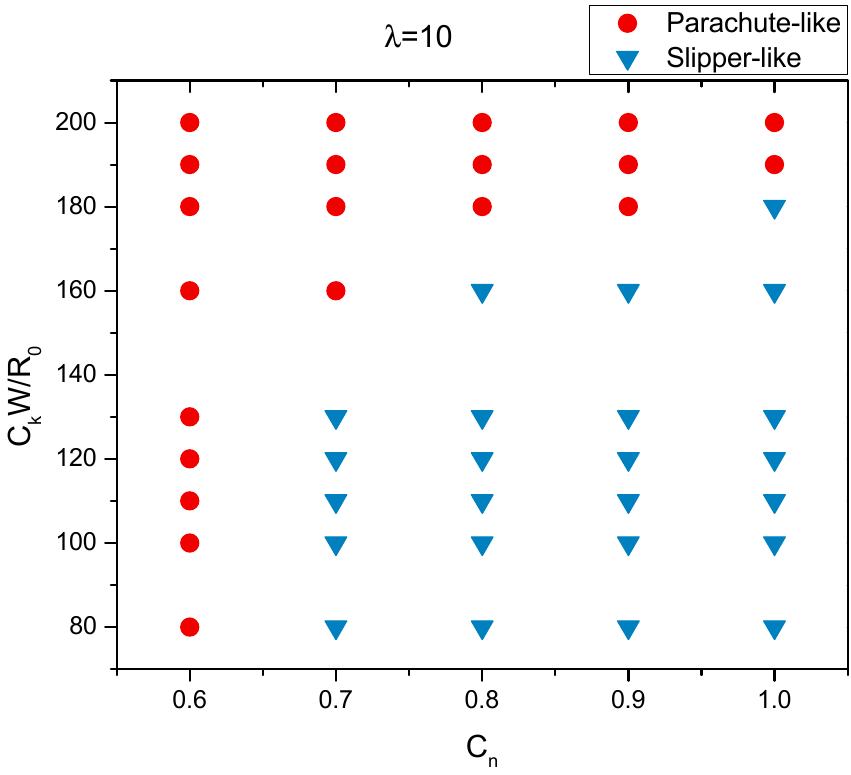}  	
   \end{minipage}
    \caption{(Color online) Phase diagram for $\lambda=10$. The red and blue dots code respectively for parachutelike and slipperlike shapes.}
   \label{fig::PhDiL10}
 \end{figure}

The complex dynamics discussed above occur at low enough flow strength. We will examine now what happens at large enough flow strength by exploring other viscosity contrasts. We will start our study by fixing the viscosity ratio to $\lambda = 5$ ($\equiv$ a cytoplasmic viscosity of around \unit[5]{cP}), which corresponds to the one of a young red cell. %
Recently Tahiri et al.~\cite{Tahiri12} investigated numerically the deformation of a single vesicle bounded by two quasi-rigid walls (walls could deform slightly) using a boundary integral %
formulation in two dimensions. They reported, in addition to the symmetric and asymmetric regions, on a region of parameter space %
where there is a coexistence between the symmetric and asymmetric shapes (parachute and slipper). %
We have reinvestigated the effect of both confinement and capillary number on the morphology of a single vesicle for the case of rigid walls. %
We have observed two possible solutions for the range of parameters investigated namely: i)-parachutelike shapes and  ii)-slipperlike shapes (Fig.~\ref{fig::Shapes_L5-L10}). %
We have summarized the results in a phase diagram in (Fig.~\ref{fig::PhDiL5}). % (slipperlike). %; and iii)-a region of coexistence between symmetric and %
Similar behavior was reported experimentally and discussed in \cite{Schoenbein81}.
We restrict the use of the word parachute for the strictly symmetrical solutions, where the word slipper covers the asymmetrical solutions. %
We have found series of symmetric-asymmetric-symmetric transitions. %
This transition was also observed in the experimental work of Abkarian et al.~\cite{Abkarian08_poiseuille} and Tomaiuolo et al.~\cite{Tomaiuolo09}, but not discussed in details. %
Tahiri et al.~\cite{Tahiri12} report that a change in the inner viscosity of the cell from around \unit[1]{cP} (viscosity of the plasma), to around \unit[5]{cP}  (a typical value for a young red cell) %
leads to different stationary shapes. Given the importance of this parameter we have also investigated another larger value. %
It is important first to underline that (i) the cytoplasmic viscosity of the red cell is a variable from one cell to the other (within the same organism), due to age, and then (ii) its value depends on the mean corpuscular hemoglobin concentration (MCHC). %
The MCHC describes the concentration of the hemoglobin per unit volume of red cell. %
Cokelet and Meiselman report that the value of the cytoplasmic viscosity increases in a non-linear manner with the MCHC ~\cite{Cokelet68}. %
During its lifespan, the mean cell volume (MCV) and the mean surface area of the red blood cell decrease with a constant ratio: the reduced volume of the cell remains the same \cite{Linderkamp82, Guido09}.
Since the concentration of the hemoglobin stays constant over time, the MCHC  increases as function of the age of the cell. A typical value of the cytoplasmic viscosity for a young red blood cell %
is around \unit[5-7]{cP}, and corresponds to a value of MCHC of about $32 g/dl$ \cite{mohandas2008red,Guido09}. For MCHC around $40 g/dl$, the viscosity of the cell nearly quadruplates \cite{Chien87}. %
Therefore, one natural question is the impact of the cytoplasmic viscosity of the red cell on dynamics.
We would like to see how does the phase diagram change (at high enough flow strength, corresponding to physiological values) when the viscosity contrast is high enough as compared to the so-called normal one, $\lambda=5$. %
To fix the ideas we have set $\lambda=10$, which corresponds to a cytoplasmic viscosity of around \unit[10]{cP}. %
We report the results in Fig.~\ref{fig::PhDiL10}. %
We observe that the slipperlike solution prevails when increasing the confinement and disappears for a $C_k W/R_0 \geq 190$. %
The separation region between the symmetric and asymmetric solutions is more pronounced than for the case of $\lambda=5$. %
Indeed, for the range of the explored data, we do not observe any kind of transition from symmetry-asymmetry-symmetry (as for $\lambda = 5$), but rather a transition from symmetric to asymmetric shapes. %
We show clearly that the stationary solutions are sensitive to inner viscosity changes, as shown in Fig.~\ref{fig::Shapes_L5-L10}.
Considering that in most of the experimental works the cytoplasmic viscosity of the red cells is an unknown variable and most probably a non uniform one, %
this may give a lead about why for a fixed flow and confinement conditions, symmetric and asymmetric shapes can both be observed. %
Our study regarding this effect is only indicative and a systematic analysis should be postponed to the future.
%
%%%%%%%%%%%%%%%%%%%%%%%%%%%%%%%%%%%%%%%%%%%%%%%%%%%%%%%%%%%%%%%%%%%%%%%%%%%%%%%%%%
%%%%%%%%%%%%%%%%%%%%%%%%%%%%%%%%%%%%%%%%%%%%%%%%%%%%%%%%%%%%%%%%%%%%%%%%%%%%%%%%%%
\section{Concluding remarks}
The most pronounced result of our study is the discovery of surprisingly complex behavior of vesicles in a Poiseuille flow. The dynamics has revealed 9 major distinct shapes and dynamics, ranging from symmetric and non-symmetric solutions, up to chaos. %
Dynamics of vesicles are treated here in the Stokes regime. In the absence of inertia, it is a classical result that the Poiseuille flow is always laminar. %
The existence of a single elastic object within the flow, acting only via bending forces, completely destroys the overall picture: chaotic dynamics take place. %
It would be interesting to investigate in the future the behavior of these chaotic regimes in the presence of many vesicles. %
It is tempting to conjecture that the composite fluid would look like chaotic both in time (as reported here) but also in space.
This problem could be viewed as a class of systems exhibiting the so-called elastic turbulence \cite{Groisman00}, %
that is a turbulence caused by the elasticity of the suspending entities when coupled to fluid flow in the purely Stokes regime. %
Elastic turbulence is characterized by a cascade of transfer of energy from large to small scales, akin to the Kolmogorov cascade for classical turbulence. %
A systematic analysis should be undertaken before drawing conclusive answers. %

In a two dimensional unbounded Poiseuille flow, the shape diagram of vesicles shows centered symmetric (parachute and bullet) shapes, and off-centered asymmetric (slipper) shapes \cite{Kaoui09}. %
These results are also observed in three dimensional simulations \cite{Farutin2014}. %
Snaking oscillations are observed in both two dimensional \cite{Kaoui11} and three dimensional \cite{Fedosov2013} simulations of vesicles in a confined Poiseuille flow. %
Chaotic dynamics, which were not reported in the previous numerical studies, were observed here under a close scrutiny. %
It is likely that the kind of solutions reported in this study should also occur in three dimensions with a rigorous investigation. %
To the limits of the authors' knowledge, there are no results in two dimensions that have not been confirmed in three dimensions. %
For all these reasons, it would be desirable to extend this work in three dimensions.
% For all these reasons, we think that the finding of this study should be expected as well in three-dimensions simulations.}

% Slipper solutions \cite{Farutin2014} as well as snaking solutions \cite{Fedosov2013} are also observed in 3D simulations. %
% It is likely that the kind of solutions reported here should also occur in 3D. %
We hope that this study will trigger further investigations both numerically and experimentally. %

\begin{acknowledgments}
We acknowledge many fruitful discussions with Dr. G. Ghigliotti at University of Nice Sophia Antipolis, and %
Dr. A. Farutin at University Grenoble Alpes.
This work was supported by the German Science Foundation research initiative SFB1027 and the graduate school GRK 1276, by DFH/UFA (the German French University). M.T. and C.M. acknowledge financial support from CNES (Centre d'Etudes Spatiales) and ESA (European Scape Agency).
\end{acknowledgments}

%%%%%%%%%%%%%%%%%%%%%%%%%%%%%%%%%%%%%%%%%%%%%%%%%%%%%%%%%%%%%%%%%%%%%%%%%%%%%%%%%%
%%%%%%%%%%%%%%%%%%%%%%%%%%%%%%%%%%%%%%%%%%%%%%%%%%%%%%%%%%%%%%%%%%%%%%%%%%%%%%%%%%
\bibliographystyle{apsrev4-1}

\end{document}